\documentclass[3p, 11pt, sort&compress]{elsarticle}
\pdfoutput = 1
\usepackage{graphics}
\usepackage{amssymb,amsmath, slashed}
\usepackage{epsfig}
\usepackage{color}
\newcommand{\bea}{\begin{eqnarray}}
\newcommand{\eea}{\end{eqnarray}}

 \makeatletter
\def\alt{\mathrel{\mathpalette\gl@align<}}
\def\agt{\mathrel{\mathpalette\gl@align>}}
\def\gl@align#1#2{\lower.6ex\vbox{\baselineskip\z@skip\lineskip\z@
\ialign{$\m@th#1\hfil##\hfil$\crcr#2\crcr\sim\crcr}}} \makeatother

\begin{document}

\begin{frontmatter}
\begin{flushright}
\begin{footnotesize}
  MAN/HEP/2014/04
\\
  May 2014
\end{footnotesize} 
\end{flushright}

\vspace{-0.5cm}

\title{{\bf Direct Bounds on Electroweak Scale Pseudo-Dirac Neutrinos \\[1mm] 
from $\sqrt s=8$ TeV LHC Data}}

\author[a]{Arindam Das}
\author[b]{P.~S.~Bhupal Dev}
\author[a]{Nobuchika Okada}

\address[a]{Department of Physics and Astronomy, 
University of Alabama,  
Tuscaloosa,  Alabama 35487, USA 
} 
\address[b]{Consortium for Fundamental Physics, 
School of Physics and Astronomy, \\
University of Manchester, 
Manchester M13 9PL, United Kingdom}
\begin{abstract}
Seesaw models with a small lepton number breaking can naturally accommodate electroweak-scale pseudo-Dirac neutrinos with a sizable mixing with the active neutrinos, while satisfying the light neutrino oscillation data. Due to the smallness of the lepton number breaking parameter, the `smoking gun' collider signature of same-sign dileptons is suppressed, and the heavy neutrinos in these models would manifest at the LHC dominantly through lepton number conserving trilepton final states. 
Using the recent CMS results for anomalous production of multilepton events at $\sqrt{s}$=8 TeV LHC with an integrated luminosity of $19.5~{\rm fb}^{-1}$, we derive {\em direct} upper bounds 
on the light-heavy neutrino mixing parameter as a function of 
 the heavy Dirac neutrino mass. These limits extend the collider sensitivity to higher 
heavy neutrino masses up to about 500 GeV. In the lower mass range, our limits are comparable to the existing indirect constraints derived from Higgs and electroweak precision data.  
\end{abstract}

\medskip


\end{frontmatter}

\section{Introduction}

The existence of nonzero neutrino masses and flavor mixing
 has been unequivocally established by various neutrino oscillation 
 experiments~\cite{PDG}. A precise understanding of 
the observed smallness of neutrino masses, as compared to all other Standard Model (SM) particles, 
could 
shed light on the underlying new physics beyond the electroweak scale. 
The seesaw extension of the SM~\cite{Seesaw} 
 is arguably the simplest idea to naturally explain 
 the tiny neutrino masses. In the canonical type-I seesaw mechanism, 
there exist SM-gauge singlet heavy Majorana neutrinos $N_\alpha$ (with the flavor index $\alpha$), which couple 
to the SM lepton doublets 
$L_l\equiv (\nu_{L,l} \quad l_L)^{\sf T}$ (with $l=e,\mu,\tau$) and the Higgs doublet $\Phi$ via the Yukawa Lagrangian  
$-{\cal L}_Y = Y_{l\alpha}\bar{L}_l\Phi N_\alpha+{\rm H.c.}$ After the electroweak 
symmetry is spontaneously broken by the Higgs vacuum expectation value (VEV) 
$\langle \Phi \rangle =(0 \quad v/\sqrt{2})^{\sf T}$, this leads to the Dirac mass matrix $M_D=vY/\sqrt{2}$  
which, along with the 
Majorana mass matrix $M_N$ of the heavy neutrinos, determines the light neutrino masses 
by the seesaw formula~\cite{Seesaw}: $M_\nu\simeq -M_DM_N^{-1}M_D^{\sf T}$.

In the minimal seesaw extension of the SM, the Majorana mass matrix $M_N$ could be arbitrary, unless dictated by some symmetry, 
as long as it reproduces the observed sub-eV scale of light neutrino masses. 
Thus, in principle, the heavy neutrinos could be around the electroweak scale accessible to the laboratory experiments. However, in the canonical seesaw, $M_N\sim 100$ GeV implies the Yukawa 
couplings $Y_{l\alpha} \lesssim 10^{-6}$, which are unobservable in foreseeable experiments.  
The situation can be improved if the mass matrices $M_D$ and $M_N$ have specific textures, 
which can be 
enforced by some symmetries (see e.g.~\cite{cancel}), so that a large light-heavy neutrino mixing is 
allowed even for a low seesaw scale, while satisfying the neutrino oscillation data.  
In this special case, electroweak-scale heavy Majorana neutrinos could be produced 
on-shell at the LHC with an observable cross section~\cite{collider1, Dev:2013wba}, and its subsequent decay leads to the characteristic signal of same-sign dilepton plus two jets ($l^\pm l^\pm jj$) which is being searched for at the LHC~\cite{LHC-LL}. Going beyond the minimal scenario by extending the gauge sector could further enhance 
the experimental prospects of testing the type-I seesaw mechanism~\cite{KS}.     

Since any number of singlets can be added to a gauge theory without introducing 
anomalies, one could exploit this freedom to find a natural alternative low-scale 
realization of the seesaw mechanism. The simplest such scenario is the inverse 
seesaw~\cite{InvSeesaw}, where one introduces two sets of SM-singlet fermions 
$N_{R,\alpha}$ and $S_{L,\beta}$ carrying opposite lepton numbers, i.e. $L(N_R)=+1=-L(S_L)$. 
The relevant Yukawa Lagrangian is given by 
\bea
-\mathcal{L}_{Y} \ = \ Y_{l\alpha} \bar{L}_l \Phi N_{R,\alpha} + M_{N,\alpha\beta} 
\bar{S}_{L,\alpha} N_{R,\beta} + \frac{1}{2}\mu_{\alpha\beta}\bar{S}_{L,\alpha}S^C_{L,\beta} 
+{\rm H.c.} \; ,
\label{eq1}
\eea 
where $S_L^C\equiv S_L^{\sf T}C^{-1}$ denotes the charge conjugate of $S_L$. Note that in 
(\ref{eq1}), $M_N$ is a Dirac mass term, while $\mu$ is the only Majorana mass term. 
After the electroweak symmetry breaking, the Lagrangian~(\ref{eq1}) gives rise to the   
full neutrino mass matrix in the flavor basis 
$\{\nu^C_{L,l}\: ,N_{R,\alpha}\: ,S^C_{L,\beta}\}$, as follows:  
\bea
{\cal M}_\nu \ = \ \begin{pmatrix}
{\mathbf 0} & M_D & {\mathbf 0} \\
M_D^{\sf T} & {\mathbf 0} & M_N^{\sf T} \\
{\mathbf 0} & M_N & \mu
\end{pmatrix} \ \equiv \ \begin{pmatrix} 
{\mathbf 0} & {\cal M}_D \\
{\cal M}_D^{\sf T} & {\cal M}_N
\end{pmatrix}
\label{mm} \; ,
\eea
where ${\cal M}_D=(M_D,{\bf 0})$ and ${\cal M}_N = \begin{pmatrix} {\bf 0} & M_N^{\sf T} \\ M_N & \mu 
\end{pmatrix}$. Diagonalizing (\ref{mm}) leads to the light neutrino mass matrix of the form
\bea
M_\nu  \ = \ M_DM_N^{-1}\mu \: M_N^{-1^{\sf T}} M_D^{\sf T} + {\cal O}(\mu^3) \; ,
\label{mnu}
\eea
whereas the heavy neutrinos form quasi-Dirac pairs $(N_i, \bar{N}_i)$ with masses roughly given by the eigenvalues of $M_N\mp \mu/2$. It is important to note that the smallness of the light neutrino masses is guaranteed 
by the smallness of 
$\mu$, irrespective of the Dirac masses $M_D$ and $M_N$. In the limit $\mu\to {\mathbf 0}$, 
lepton number symmetry is restored, and the light neutrinos are exactly massless. 
In other words, the 
smallness of $\mu$ is technically natural in the 't Hooft sense. Such a small mass term may be generated from some other new physics, e.g. spontaneous breaking of lepton 
number~\cite{majoron}, radiative corrections~\cite{radiative} or extra dimensions~\cite{extradim}. Similarly, the Dirac mass matrix $M_N$ and the inverse seesaw structure in (\ref{mm}) could be explained in various extensions of the minimal inverse seesaw model~\cite{Dev:2009aw, GOS, Khalil:2010iu}. Note that a nonzero Majorana mass term $\mu_R \bar{N}_R N_R^C$ could still 
be allowed in (\ref{eq1}) if $N_R$ is a gauge singlet. 
This will contribute to the light neutrino mass matrix 
in (\ref{mnu}) at one-loop level~\cite{Dev:2012sg} from standard electroweak radiative 
corrections~\cite{zpc}. In this case, both $\mu$ and $\mu_R$ contributions can be 
combined to define an effective Majorana mass $\mu_{\rm eff}$, while keeping the remaining structure in (\ref{mnu}) unchanged. 

The general feature of the inverse seesaw mechanism, i.e. a small lepton number breaking, allows large neutrino Yukawa couplings $Y_{l\alpha}$ up to ${\cal O}(1)$ even for an electroweak scale heavy neutrino mass $M_N$, without introducing any fine-tuning or cancellations in the light neutrino mass matrix (\ref{mnu}). This leads to a number of interesting phenomenological consequences, such as large lepton flavor violation (LFV)~\cite{LFV}, non-unitarity of the leptonic mixing matrix~\cite{Dev:2009aw, nuty}, light DM candidate~\cite{DM} and modifications to the SM Higgs observables~\cite{GOS, DFM, Higgs}. In this Letter, we will mostly focus on the collider signatures of this low-scale seesaw mechanism.  
\section{Trilepton Signature at the LHC}
\begin{figure}[t!]
\centering
\includegraphics[width=7cm]{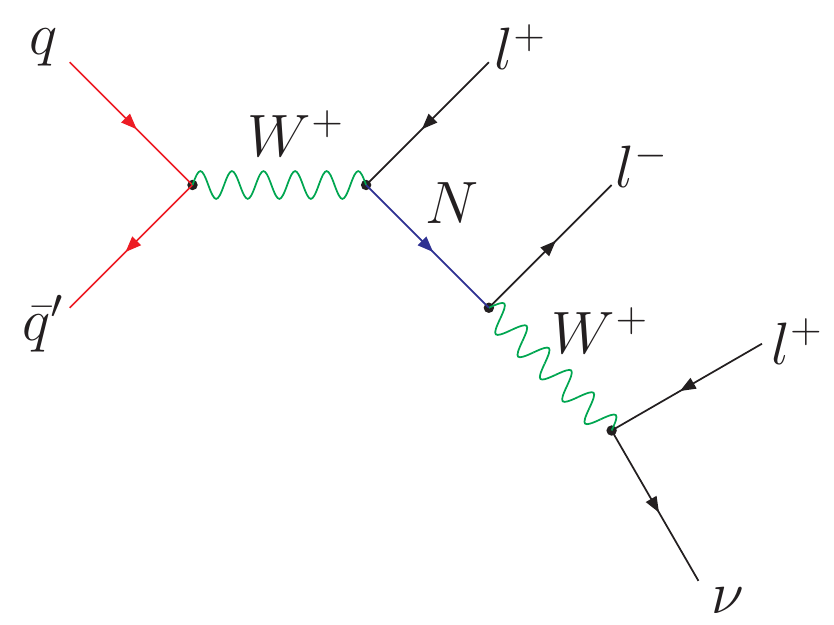}
\caption{The trilepton plus missing transverse energy signal of a heavy Dirac neutrino at the LHC.}
\label{fig1}
\end{figure}

As far as the direct collider tests of the inverse seesaw mechanism are concerned, a large Yukawa coupling enhances the on-shell production of electroweak-scale heavy neutrinos at the LHC. However, due to the small lepton number breaking in these scenarios, the heavy neutrinos are pseudo-Dirac, and hence, the `smoking gun' collider signature of same-sign dilepton final states is suppressed. As the opposite-sign dilepton signal $l^\pm l^\mp jj$ is swamped with a large SM background, mainly from 
$pp\to Zjj$, the `golden' channel for probing heavy Dirac neutrinos at the LHC is the trilepton final state~\cite{delAguila}:
\bea
pp & \to & l_1^+ N \ \to \ l_1^+ l_2^- W^+ \ \to \ l_1^+ l_2^- l_3^+ \nu \; , \nonumber \\
pp & \to & l_1^- \bar{N} \ \to \ l_1^- l_2^+ W^- \ \to \ l_1^- l_2^+ l_3^- \bar{\nu} \; ,
\label{tl}
\eea 
as shown in Figure~\ref{fig1}.\footnote{In a supersymmetric version of the inverse seesaw model, the same trilepton plus missing transverse energy final state as in (\ref{tl}) can also be obtained from a pair-production of charginos and neutralinos~\cite{trilep-SUSY}.} 
Here $N$ denotes a generic heavy neutrino mass eigenstate, which is typically the lightest 
SM-singlet fermion in a given seesaw model. From (\ref{mm}), we see that for $\|\mu\|\ll \|M_N\|$, the heavy neutrino masses given by the eigenvalues of ${\cal M}_N$ can be approximated by degenerate pairs of the eigenvalues of $M_N$. The small mass splitting between the quasi-Dirac pairs induced by the small lepton number breaking parameter $\mu$ is irrelevant for their collider studies, as long as $\mu$ is much smaller than their decay widths. This is a valid approximation in our case since we require relatively large neutrino Yukawa couplings in order to have a sizable collider signal and a very small $\mu$ to satisfy the neutrino oscillation data [cf.~(\ref{mnu})].  Thus, we can treat the heavy neutrinos to be Dirac particles for our subsequent collider analysis. It is important to note here that the trilepton signal does not vanish in the $\mu\to 0$ limit. This is in contrast with the collider signature of heavy Majorana neutrinos  in the minimal setup, where the same-sign dilepton signal must vanish in the limit of exact degeneracy.

The discovery potential of the trilepton channel (\ref{tl}) at the LHC, along with a detailed SM background analysis, was performed in~\cite{delAguila} for a single-flavor electroweak-scale heavy neutrino. A similar study in the context of Left-Right symmetric theory was presented in~\cite{ChenDev}. A more general heavy neutrino flavor structure was considered in~\cite{DasOkada}, and it was shown that a $5\sigma$ statistical significance of the signal events over the SM background can be achieved at $\sqrt s=14$ TeV LHC with 11 fb$^{-1}$ luminosity in the flavor-diagonal case.  Similar sensitivities can also be achieved at the planned ILC with $\sqrt s=500$ GeV -- 1 TeV. Note that, at an electron-positron collider, the dominant heavy neutrino production channel is $e^+e^-\to \nu N$, which leads to one isolated lepton and two jets with large missing energy signature~\cite{Achard:2001qv, DasOkada}, irrespective of the Dirac/Majorana nature of the heavy neutrinos.

Meanwhile, the CMS collaboration has presented a model-independent search for anomalous production of events with at least three isolated charged leptons using the $19.5~{\rm fb}^{-1}$ data at 
$\sqrt s=8$ TeV LHC~\cite{CMS1}. They have adopted a general search strategy which is applicable 
to  a wide range of possible scenarios beyond the SM giving rise to a multilepton signal, including the pseudo-Dirac neutrino case discussed above. With this observation, we perform a collider 
simulation of 
the trilepton signal for a generic pseudo-Dirac heavy neutrino scenario with the same selection criteria as used in the CMS analysis, and compare our signal events with the CMS observed data. Using the fact that the experimental results are consistent with the SM expectations, we derive the first {\it direct} limits on the pseudo-Dirac heavy neutrino mass and mixing with the light    neutrinos. 

The inclusive cross section for the trilepton final state given by (\ref{tl}) in a generic seesaw model can be written as 
\bea
\sigma(pp\to l_1l_2l_3+\slashed{E}_T) \ = \ \sigma_{\rm prod}(pp\to W^*\to N l_1)\:{\rm BR}(N\to l_2 W)\: {\rm BR}(W \to l_3 \nu) \; .
\label{cross}
\eea
Here we assume that the heavy neutrinos are heavier than the $W$-boson so that the two-body decay $N\to lW$ is kinematically allowed, followed by an on-shell $W$-decay to SM leptons. In general, the final state charged leptons $l_{1,2,3}$ 
can be of any flavor combination. However, since it is rather challenging to reconstruct the $\tau$-lepton events with our simulation tools, we will only consider the electron and muon final states, i.e. $l=e,\mu$. In this case, the SM $W$ branching ratio is given by BR$(W\to l\nu)=0.21$~\cite{PDG}, whereas the production cross section as well as the partial decay widths of the heavy neutrino depend on the light-heavy neutrino mixing parameter(s). 

To parametrize the light-heavy neutrino mixing, we first diagonalize the full neutrino mass matrix (\ref{mm}) by a unitary mixing matrix: 
\bea
{\cal V}^{\sf T}{\cal M}_\nu {\cal V} \ = \  {\rm diag}(m_i\: , M_j) \; ,
\label{diag}
\eea 
where $m_i$ (with $i=1,2,3$) and $M_j$ (with $j=4,5,6,...$) are respectively the light and heavy neutrino mass eigenvalues. The unitary matrix ${\cal V}$ has an exact representation in terms of 
a dimensionless matrix $\xi$ (which depends on $M_D$ and $M_N$), as follows~\cite{Dev:2012sg, Korner:1992zk}: 
\bea
{\cal V} = \left(\begin{array}{cc}
({\bf 1}+\xi^*\xi^{\sf T})^{-1/2} & \xi^*({\bf 1}+\xi^{\sf T}\xi^*)^{-1/2} \\
-\xi^{\sf T}({\bf 1}+\xi^*\xi^{\sf T})^{-1/2} & ({\bf 1}+\xi^{\sf T}\xi^*)^{-1/2}
\end{array}  \right)
\left(\begin{array}{cc}
U & {\bf 0} \\
{\bf 0} & V
\end{array}\right) \; ,
\label{V}
\eea
where $U$, $V$ are the  unitary matrices diagonalizing the light and heavy neutrino mass matrices $M_\nu$ and ${\cal M}_N$ respectively. Now using (\ref{diag}) and (\ref{V}), the light neutrino flavor eigenstates can be related to the mass eigenstates $\widehat{\nu}_i$ and $\widehat{N}_j$ as follows: 
\bea
\nu_l \ = \ \left[({\bf 1}+\xi^*\xi^{\sf T})^{-1/2}\right]_{lm} U_{mi} \widehat{\nu}_i + \left[\xi^*({\bf 1}+\xi^{\sf T}\xi^*)^{-1/2}\right]_{lk}V_{kj}\widehat{N}_j 
\ \equiv \ {\cal N}_{li} \widehat{\nu}_i + {\cal R}_{lj} \widehat{N}_j\; ,
\eea
where the first term on the right-hand side (RHS) measures the non-unitarity of the PMNS 
mixing matrix~\cite{Constraints1}, 
and the second term determines the size of the light-heavy neutrino mixing in 
charged and neutral-current interactions.  In the charged-lepton mass diagonal basis, the charged-current interaction relevant for the production and decay of heavy neutrinos at the LHC is given by 
\bea 
-\mathcal{L}_{CC} \  = \ \frac{g}{\sqrt{2}} W^-_{\mu}\bar{l} \gamma^{\mu} P_L \nu_l + {\rm H.c.} 
\  = \ \frac{g}{\sqrt{2}} W^-_{\mu}\bar{l} \gamma^{\mu} P_L  \left( {\cal N}_{li} \widehat{\nu_i}+ {\cal R}_{lj} \widehat{N}_j \right) + {\rm H.c.} \; ,
\label{CC}
\eea
where $P_L = (1- \gamma^5)/2$ is the left-chirality projection operator. Similarly, the neutral-current interaction is given by 
\bea 
-\mathcal{L}_{NC} & = &  \frac{g}{2 \cos\theta_w}  Z_{\mu} \bar{\nu}_l \gamma^\mu P_L \nu_l 
\ = \  \frac{g}{2 \cos\theta_w}  Z_{\mu} \left[ 
({\cal N}^\dag {\cal N})_{ij}  \widehat{\bar{\nu}}_i \gamma^{\mu} P_L \widehat{\nu}_j 
 +  ({\cal R}^\dag {\cal R})_{ij}\widehat{\bar{N}}_i \gamma^{\mu} P_L \widehat{N}_j \right. \nonumber \\ 
&  & 
\qquad \qquad \qquad \qquad \qquad \qquad \qquad \qquad \qquad \left.  
+ \left\{({\cal N}^\dag {\cal R})_{ij}\widehat{\bar{\nu}}_i \gamma^{\mu} P_L \widehat{N}_j 
  + {\rm H.c.} \right\} 
\right] , 
\label{NC}
\eea
 where $\theta_w$ is the weak mixing angle. From (\ref{CC}), we see that the heavy 
neutrino production cross section in (\ref{cross}) will be proportional to the mixing parameter 
$|{\cal R}_{lj}|^2$. We should note here that although our discussion so far is based on the inverse seesaw mass matrix given by (\ref{mm}), the collider analysis and the results derived subsequently are more general, and should apply to any heavy neutrino mixing with the active neutrinos, as e.g. in various low-scale seesaw models (for a recent review, see~\cite{Boucenna:2014zba}).

After being produced on-shell, the heavy Dirac neutrinos have 
the following two-body decays to the SM final states $l^- W^+$, $Z\nu$ and $h\nu$  (if kinematically allowed). The corresponding partial decay widths are given by  
\bea
\Gamma(N_i\to \ell^- W) & = & \frac{g^2}{64\pi}|{\cal R}_{l i}|^2\frac{M_i^3}{M_W^2}\left(1-\frac{M_W^2}{M_i^2}\right)^2\left(1+2\frac{M_W^2}{M_i^2}\right),\\
\Gamma(N_i\to \nu_\ell Z) & = & \frac{g^2}{128\pi\cos^2\theta_w}|{\cal R}_{li}|^2\frac{M_i^3}{M_Z^2}\left(1-\frac{M_Z^2}{M_i^2}\right)^2
\left(1+2\frac{M_Z^2}{M_i^2}\right),\\
\Gamma(N_i\to \nu_\ell h) & = & \frac{g^2}{128\pi}|{\cal R}_{li}|^2\frac{M_i^3}{M_W^2}\left(1-\frac{M_h^2}{M_i^2}\right)^2. 
\label{widths}
\eea 
Using these expressions, one can calculate the branching ratio of the decay $N\to l^-W^+$ in (\ref{cross}). Note that BR($N\to l^-W^+$) in the heavy Dirac neutrino case is twice 
larger as compared to the heavy Majorana neutrino case, which has equal probability to decay into either $l^-W^+$ or $l^+W^-$. 

\section{Benchmark Scenarios}

The flavor information of the final state leptons $l_1^\pm l_2^\mp l_3^\pm$ due to the production and decay of a particular heavy neutrino mass eigenstate $N_i$ can be parametrized by 
$|{\cal R}_{l_1 i}{\cal R}_{l_2 i}|^2/\sum_{l=e,\mu} |{\cal R}_{li}|^2$. For the collider simulation of our trilepton signal (\ref{tl}), we make some reasonable simplifying assumptions. First of all, to leading order in a converging Taylor series expansion of $\xi$, the mixing matrix ${\cal R}$ can be approximated by ${\cal R}\simeq \xi^* = M_DM_N^{-1}$. Furthermore, we assume flavor-diagonal Dirac mass matrices $M_D$ and $M_N$, which suppress all LFV processes. This will always lead to a trilepton final state with opposite sign same flavor charged leptons (OSSF1 in the notation of~\cite{CMS1}). Using the CMS observed number of events and the corresponding SM expectation values for the OSSF1 case~\cite{CMS1}, we will obtain direct constraints on the diagonal mixing elements of ${\cal R}$ in a model-independent way. Note that the light neutrino oscillation data can still be satisfied with flavor-diagonal $M_D$ and $M_N$ 
by a suitable flavor structure of the lepton number breaking parameter $\mu$ in (\ref{mnu}); see~\cite{DasOkada} for an explicit numerical fit. For illustration, we will consider the following two benchmark cases with different relative magnitudes between the flavor Yukawa couplings: 
\begin{itemize}
\item [(a)] Single flavor (SF) case, in which one heavy Dirac pair is at the electroweak scale, while other heavy pairs are assumed to be beyond the kinematic reach of the LHC. With our flavor-diagonal choice, the lightest heavy Dirac neutrino mass eigenstates dominantly couple to a single lepton flavor $l$, which we assume to be muon for concreteness, although the same study equally applies for the electron-flavor case. Thus, we have the trilepton final states $l_1^\pm l_2^\mp l_3^\pm$ with both $l_1$ and $l_2$ of the muon flavor, while 
$l_3$ coming from the $W$-decay can be either electron or muon, i.e. our final state flavor compositions are $\mu^\pm \mu^\mp e^\pm$ and $\mu^\pm \mu^\mp \mu^\pm$. Note that in this case, the lightest heavy-neutrino branching fraction in 
(\ref{cross}) is independent of the mixing ${\cal R}_{\mu 1}$, and the only dependence on the 
mixing parameter appears in the production cross section, which is proportional to $|{\cal R}_{\mu 1}|^2$.  Thus, we can derive constraints on the single-flavor mixing parameter $|{\cal R}_{\mu 1}|^2 \equiv |B_{\mu N}|^2$, as a function of the lightest heavy Dirac neutrino mass $m_N$.

\item [(b)] Flavor diagonal (FD) case, in which two of the heavy Dirac neutrino pairs are degenerate with a common mass $m_N$. We further assume that one pair dominantly couples 
to electrons, and the other one to muons, but with equal strength, i.e., $|{\cal R}_{e 1}| = |{\cal R}_{\mu 2}| \equiv |B_{lN}|$.\footnote{Note that in the inverse seesaw with pseudo-Dirac neutrinos, the lepton number violating process of neutrinoless double beta decay ($0\nu\beta\beta$) is usually  suppressed, and therefore, the  stringent constraints on the active-sterile neutrino mixing in the electron sector (see e.g.,~\cite{Atre}) may not apply. Hence, our benchmark case (b) with relatively large $|B_{eN}|^2$ is still allowed, except in special cases where the $0\nu\beta\beta$ amplitude could be enhanced~\cite{0vbb}.} In this case, we have two additional final states $e^\pm e^\mp e^\pm$ and $e^\pm e^\mp \mu^\mp$, along with the two final states of the benchmark case (a). Thus the total trilepton signal cross section in case (b) is twice larger than that in case (a), and as a result, the limit on  $|B_{l N}|^2$ derived in case (b) will be roughly twice stronger than the corresponding limit in case (a) for a given value of $m_N$. 
\end{itemize} 

\section{Data Analysis and Results}
For each of the above benchmark cases, the trilepton signal events were generated for $\sqrt s=8$ TeV LHC by implementing the new interaction vertices given by (\ref{CC}) and (\ref{NC}) in {\tt MadGraph5}~\cite{Mad}. The parton level cross sections were obtained using the {\tt CTEQ6L} parton distribution functions~\cite{CTEQ}. The showering and hadronization of the events were performed with {\tt PYTHIA6.4}~\cite{Pythia} and a fast detector simulation was done using {\tt DELPHES3}~\cite{Delp}. Hadrons were clustered into jets using the anti-$k_T$ algorithm as implemented in {\tt FastJet2}~\cite{fastjet} with a distance parameter of 0.5. In the detector simulation, we have considered the signal events containing three leptons accompanied by $n$-jets (with $n=1$--4), after incorporating the MLM matching prescription~\cite{MLM} to avoid double counting of jets. For the generated signal events, we adopt the following basic selection criteria, as used in the CMS trilepton analysis~\cite{CMS1}:
\begin{itemize}
\item [(i)] The transverse momentum of each lepton: $p^l_T > 10$ GeV.
\item [(ii)] The transverse momentum of at least one lepton: $p^{l,{\rm leading}}_{T} > 20$ GeV.
\item [(iii)]  The jet transverse momentum: $p_T^j > 30$ GeV. 
\item [(iv)] The pseudo-rapidity of leptons: $|\eta^l| < 2.4$ and of jets: $|\eta^j| < 2.5$.
\item [(v)] The lepton-lepton separation: $\Delta R_{ll} > 0.1$ and the lepton-jet separation: $\Delta R_{lj} > 0.3$. 
\item [(vi)] The invariant mass of each OSSF lepton pair: $m_{l^+ l^-}< 75$  GeV or $> 105$ GeV to avoid the on-$Z$ region which was excluded from the CMS search. Events with $m_{l^+ l^-}< 12$ GeV are rejected to eliminate background from low-mass Drell-Yan processes and hadronic decays. 
\item [(vii)]  The scalar sum of the jet transverse momenta: $H_{T} < 200$ GeV. 
\item [(viii)] The missing transverse energy: $\slashed{E}_{T} < 50$ GeV. 
\end{itemize}
Note that there are additional contributions to the trilepton signal from $N\to Z\nu, h\nu$, followed by $Z, h$ decay to $l^+l^-$. However, the $Z$ contributions are suppressed after we impose the $m_{ll}$ cut to reduce the SM $Z$ background, whereas the $h$ contributions are additionally suppressed due to small Yukawa coupling of electrons and muons. 
  The CMS analysis~\cite{CMS1} has given the number of 
observed events and the corresponding SM background expectation for various ranges of $\slashed{E}_T$ and $H_T$ that are sensitive to different kinematical and topological signatures. However, for our trilepton signal (\ref{tl}), the set of selection cuts listed above turn out to be the most efficient ones among those considered in the CMS analysis. 

\begin{figure}[t!]
\centering
\includegraphics[width=7cm]{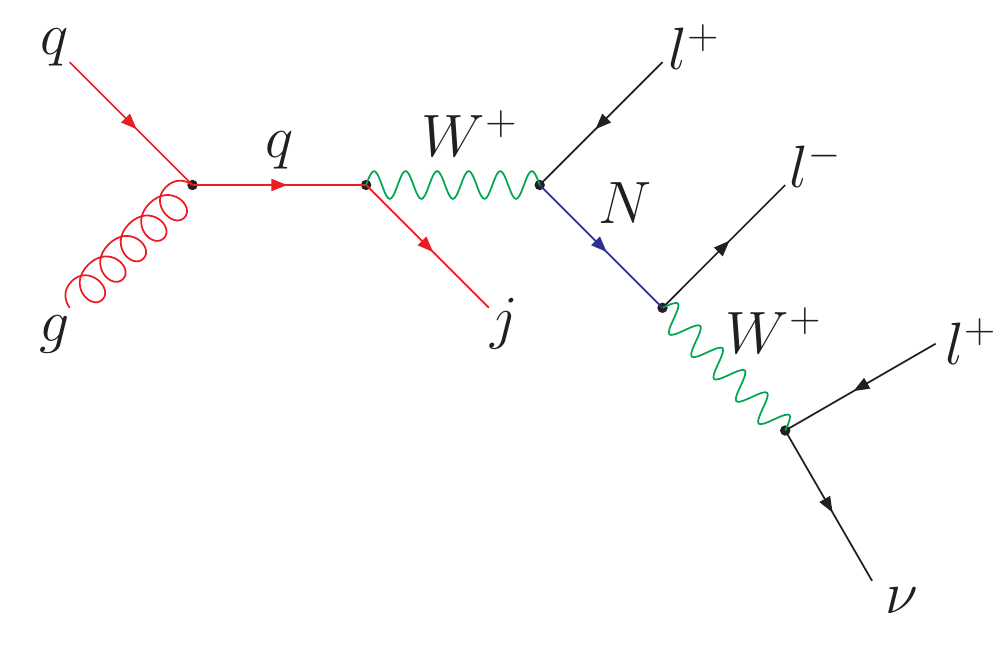}
\includegraphics[width=7cm]{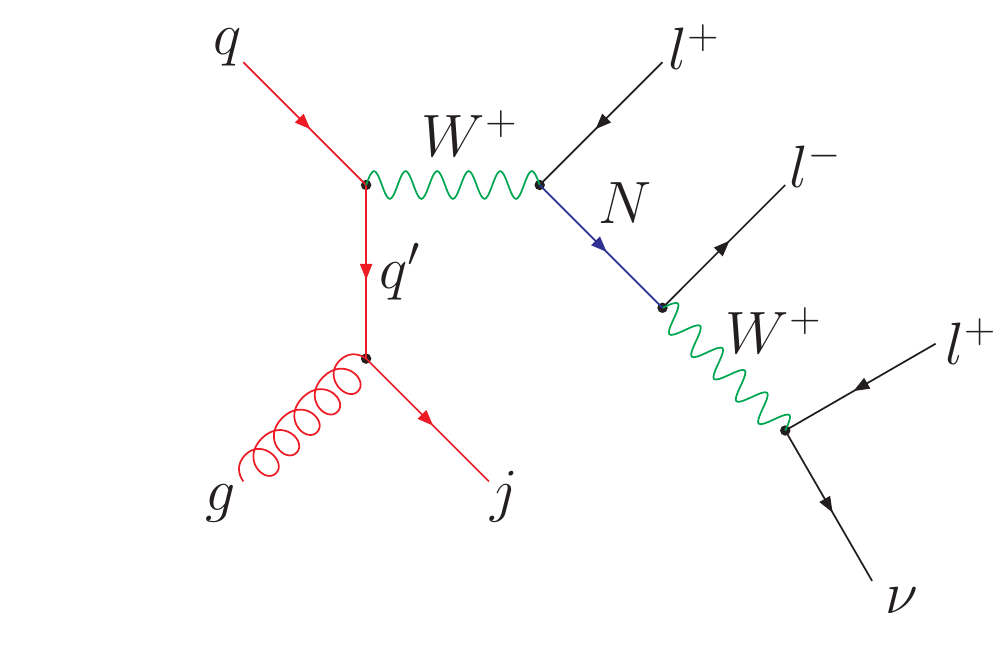}
\caption{The trilepton + one jet + missing transverse energy signal of a heavy Dirac neutrino at the LHC.} 
\label{fig2}
\end{figure}

\begin{figure}[t!]
\centering
\includegraphics[width=10cm]{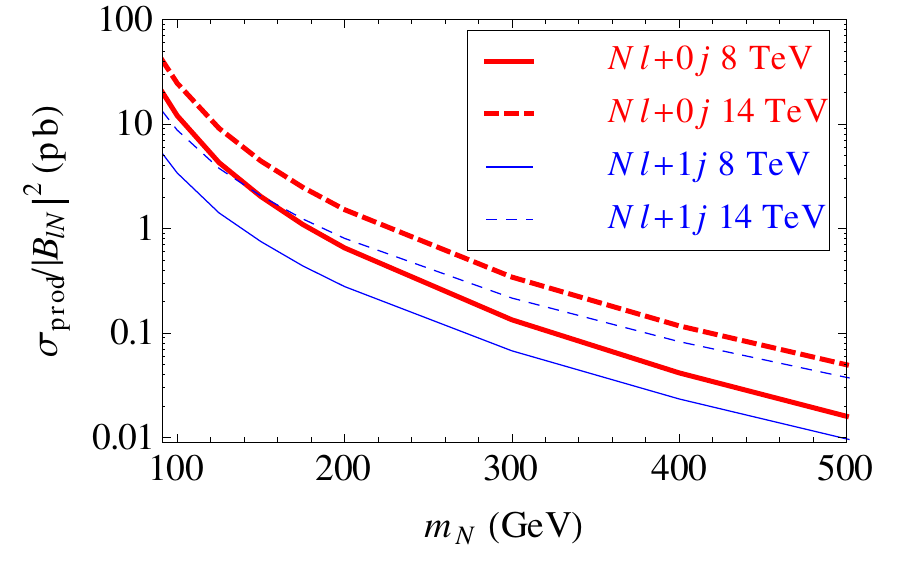}
\caption{The `inclusive' parton-level cross sections for the processes $pp\to Nl^++\bar{N}l^-$ (thick, red) and $pp\to Nl^+j+\bar{N}l^-j$ (thin, blue) at $\sqrt s=8$ TeV (solid) and 14 TeV (dashed) LHC. 
The results are shown for the single flavor (SF) case. For the flavor diagonal (FD) case, the numbers should be multiplied by a factor of two. For the $Nlj$ case, we have imposed $p_T^j>30$ GeV.}
\label{fig3}
\end{figure}

It is important to note here that in order to make a direct comparison of our signal events with the 
CMS results for the observed events and the SM background, we must include at least one jet with $p_T>30$ GeV and $|\eta^j|<2.5$ in the 
final state. The simplest trilepton final state shown in Figure~\ref{fig1} does not contain any jets at the parton-level, but initial state radiation (ISR) effects could give rise to final states with non-zero jets, though they are usually expected to be soft. However, there are additional diagrams involving quark-gluon fusion, such as those shown in Figure~\ref{fig2}, which give rise to hard jets in the final state. The inclusive production cross section of the processes $pp\to Nl^+ (\bar{N}l^-)+1j$ is only a factor of 2--4 smaller than the original $pp\to Nl^+ (\bar{N}l^-)+0j$ process shown in Figure~\ref{fig1}. This is due to the fact that, although the three-body final state $Nlj$ is phase-space suppressed compared to the two-body final state $Nl$, there is a partially compensating enhancement at the LHC due to a much larger gluon content of the proton, as compared to the quark content~\cite{PDG}. The numerical values of the two production cross sections, normalized to $|B_{lN}|^2=1$,  are shown in Figure~\ref{fig3} for both 
$\sqrt s=8$ and 14 TeV LHC as a function of the lightest heavy neutrino mass $m_N$. Here we have shown the values for the SF case; for the FD case, the cross sections are enhanced by a factor of two. Note that for the $Nl+1j$ case, we must use a non-zero $p_T^j$ cut to avoid the infrared singularity due to massless quarks in the $t$-channel. 
Here we have used the $p_T^j>30$ GeV cut, following the CMS analysis, to get a finite result. Using a lower value of $p_T^{j,{\rm min}}$ could enhance the $Nl+1j$ cross section, 
thereby improving the signal sensitivity. Moreover, for a lower $p_T^{j,{\rm min}}$, other processes such as $pp\to Nljj$ mediated by a $t$-channel photon exchange~\cite{Dev:2013wba} and $gg\to Nljj$  mediated by $t$-channel quarks, could give additional enhancement effects. A detailed detector-level simulation of these infrared-enhanced  processes for different selection criteria than those used by the current CMS analysis is beyond the scope of this Letter, and will be presented in a separate communication. In this sense, the bounds on light-heavy neutrino mixing derived here can be treated as conservative bounds.    

\begin{figure}[t!]
\begin{center}
\includegraphics[width=10cm]{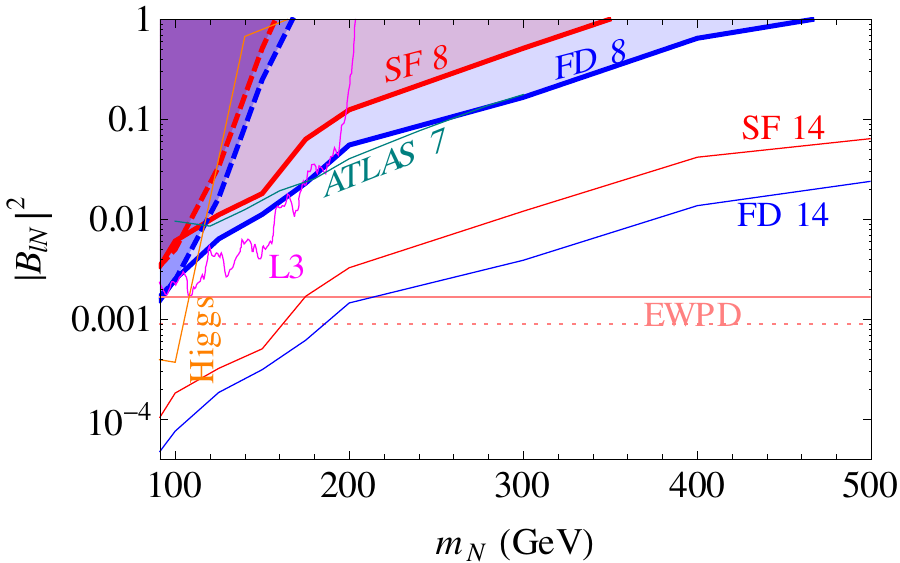}
\end{center}
\caption{The 95\% CL upper limit on the light-heavy neutrino mixing parameter $|B_{lN}|^2$ as 
a function of the heavy Dirac neutrino mass $m_N$, derived from the CMS  
trilepton data at $\sqrt s=8$ TeV LHC for $19.5~{\rm fb}^{-1}$ luminosity~\cite{CMS1}. 
The exclusion (shaded) regions are shown 
for two benchmark scenarios: (i) single flavor (SF) and (ii) flavor diagonal 
(FD), with two choices of the selection cut $m_{l^+l^-}<75$ GeV (thick dotted) and $>105$ GeV (thick solid). The corresponding conservative projected limits from $\sqrt s=14$ TeV LHC data with $300~{\rm fb}^{-1}$ integrated luminosity are shown by thin solid lines (SF 14 and FD 14). Some relevant existing upper limits (all at 95\% CL) are also shown for comparison: (i) from a $\chi^2$-fit to the LHC Higgs data~\cite{DFM} (Higgs), (ii) from a direct search at LEP~\cite{Achard:2001qv} (L3), valid only for the electron flavor, (iii) ATLAS limit from $\sqrt s=7$ TeV LHC data~\cite{LHC-LL} (ATLAS 7), valid for a heavy {\it Majorana} neutrino of the muon flavor, and (iv) indirect limit from a global fit to the electroweak precision data~\cite{EWPD} (EWPD), for both electron (solid) and muon (dotted) flavors. } 
\label{fig4}
\end{figure}

To derive the limits on $|B_{lN}|^2$, we calculate the normalized signal cross section $\sigma/|B_{lN}|^2$ at $\sqrt s=8$ TeV LHC as a function of the lightest heavy neutrino mass $m_N$ for both SF and FD cases, after imposing the CMS selection criteria listed above. The corresponding number of signal events passing all the cuts is then compared with the observed number of events for 19.5~${\rm fb}^{-1}$ luminosity~\cite{CMS1}.  For the selection criteria listed above, the CMS experiment observed (a) 510 events with the SM background expectation of $560\pm 87$ events for $m_{l^+l^-}<75$ GeV and (b) 178 events with the SM background expectation of $200\pm 35$ events for $m_{l^+l^-}>105$ GeV. Thus, for case (a), we have an upper limit of 37 signal events, and for case (b) an upper limit of 13 signal events. This sets a direct upper bound on the light-heavy neutrino mixing parameter $|B_{lN}|^2$ for a given value of $m_N$, as shown in Figure~\ref{fig4} for both cases (a) and (b) discussed above (thick dashed and solid lines, respectively). The case (b) becomes more efficient for higher values of $m_N$, thus setting a more stringent limit on $|B_{lN}|^2$. We have shown the 
95\% CL exclusion regions for both benchmark scenarios, namely SF and FD cases (red and blue shaded regions, respectively). As expected, the upper bound in the FD case is roughly twice as stronger than that in the SF case.  In our analysis, we have considered heavy neutrino masses only above $M_Z$, since for $m_N<M_Z$, the existing LEP limits from $Z$-decay~\cite{LEP} are more stringent. 

For comparison, we also show the 95\% CL indirect upper limit on $|B_{lN}| < 0.030~(0.041)$ for $l=\mu~(e)$ derived from a global fit to the electroweak precision data~\cite{EWPD} (EWPD), which is independent of $m_N$ for $m_N>M_Z$, as shown by the horizontal dotted (solid) line in Figure~\ref{fig4}. We find that the direct bounds on $|B_{lN}|^2$ derived here are comparable to the indirect ones for $m_N\sim 100$ GeV, but get weaker at higher masses due to the suppression in the heavy neutrino production cross section (cf. Figure~\ref{fig3}). Similar but somewhat weaker indirect bound could also be obtained from non-unitarity of the leptonic mixing matrix and lepton flavor universality constraints~\cite{Constraints1}. In addition, 95\% CL constraints on the Yukawa coupling, and hence, on the mixing parameter $|B_{lN}|^2$ could be obtained from a $\chi^2$-fit to the LHC Higgs data~\cite{DFM}, as shown by the orange solid line (Higgs) in Figure~\ref{fig4}. This limit turns out to be the strongest one for $m_N\lesssim M_h$, but becomes ineffective for larger $m_N$ as $N$ becomes off-shell in the Higgs decay $h\to N\nu \to 2l2\nu$. 

Finally, we also compare the direct limits derived here with the existing collider bounds. The 95\% CL LEP limit on $|B_{eN}|^2$, derived from the search channel $e^+e^-\to N_e\nu_e \to eW\nu_e$~\cite{Achard:2001qv}, is shown by the pink solid line (L3) in Figure~\ref{fig4}. For a small range of the parameter space, this limit is stronger than the LHC trilepton limit derived here. However, the LEP limit is only applicable to the electron flavor, whereas the trilepton limit derived here is equally applicable to both electron and muon flavors. Moreover, the trilepton final states are also applicable to the heavy Majorana neutrino case, although the smoking gun collider signature in the Majorana case will be the same-sign dilepton final state, which is dominant over the trilepton signal. For completeness, we have shown the corresponding limits from a same-sign dimuon search by ATLAS using the $\sqrt s=7$ TeV LHC data for $4.7~{\rm fb}^{-1}$~\cite{LHC-LL}.  These limits are comparable to the trilepton limits derived here; however, a dedicated search optimized for the trilepton signal could lead to a more stringent limit than that presented here. 

In light of the above results and the noncompetitiveness of the direct bounds obtained here with the existing indirect limits, it might be useful to derive the projected direct limits anticipated from the $\sqrt s=14$ TeV LHC data. Assuming that the signal efficiency is the same as that obtained for the $\sqrt s=8$ TeV data analysis with $m_{l^+l^-}>105$ GeV selection cut, and using the production cross sections given in Figure~\ref{fig3}, we obtain the projected upper limits on $|B_{lN}|^2$ for both SF and FD cases at the $\sqrt s=14$ TeV LHC with $300~{\rm fb}^{-1}$ integrated luminosity, as shown by the thin solid red (SF 14) and blue (FD 14) lines in Figure~\ref{fig4}.  These limits should be treated as conservative limits, since the signal-to-background selection efficiency at $\sqrt s=14$ TeV LHC is expected to be {\it at least} as good as that in the $\sqrt s=8$ TeV case. Thus, we find that the direct limits on the heavy-light neutrino mixing parameter are expected to improve significantly (by at least one order of magnitude) at the $\sqrt s=14$ TeV LHC.
\section{Conclusion}
In this Letter, we have derived the first direct collider bounds on electroweak-scale pseudo-Dirac heavy neutrinos, which could be naturally motivated in inverse seesaw models to explain the observed smallness of active neutrino masses by a small lepton number breaking. The derived upper bound on the light-heavy neutrino mixing parameter $|B_{lN}|^2$ is about $2\times 10^{-3}$ for $m_N\sim 100$ GeV, and is comparable to the existing best limit from electroweak precision tests. Our analysis provides the first direct limits on the mixing parameter $|B_{lN}|^2$ up to $m_N=500$ GeV or so.    
The bounds derived here should be considered as conservative bounds, since optimizing the experimental analysis for our particular trilepton channel, and including the infrared enhancement effects due to $t$-channel quarks and photons, will yield a much stronger bound. We hope the experimental community will seriously consider this possibility. Finally, the collider bounds could significantly improve with more data from the upcoming 
LHC run-II with $\sqrt s= 13-14$ TeV.


\section*{Acknowledgments}
We thank Keith Ulmer and Frank Wuerthwein for sharing useful information 
regarding the CMS data used in our analysis, and 
John Ellison for carefully reading the manuscript. A.D. also thanks Pradipta Ghosh for useful 
discussions on the numerical tools. P.S.B.D. would like to thank Apostolos Pilaftsis, Vladimir Savinov and Un-ki Yang for helpful discussions on heavy neutrino searches at the LHC. 
The work of P.S.B.D. is supported by
 the Lancaster-Manchester-Sheffield Consortium for Fundamental 
 Physics under STFC grant ST/J000418/1. The work of N.O. is supported in part by the DOE Grant No. DE-FG02-10ER41714. 

\end{document}